# Data-driven discovery of a model equation describing self-oscillations of direct current discharge

Dmitry Levko

Esgee Technologies Inc., Austin, Texas 78746, USA

E-mail: dima.levko@gmail.com

**Abstract**

Data-driven techniques developed in recent years for the discovery of equations describing complex physical phenomena open unique opportunities for plasma physics. These methods allow getting insights into the processes difficult for analytical description. Since gas discharges can be represented as complex electrical circuits consisting of impedances and capacitances, it looks natural to use the data-driven techniques to study their complex dynamics. In the present paper, the sparse identification of nonlinear dynamics (SINDy) method is applied to analyze the self-oscillations of direct current discharge in argon. It is obtained that the third order polynomials describe best the oscillations of the discharge voltage and current. They allow an accurate capturing of the oscillations amplitudes as well as the harmonics of these oscillations. To understand the physical meaning of each term, an analytical model is presented which describes the discharge self-oscillations.

## 1. Introduction

Direct current (dc) glow discharges have been widely studied since the early days of gas discharge physics [1]. They are of both fundamental and practical importance. Fundamentally, they are of interest because allow testing of various plasma models, both theoretical and computational. They can also be used to measure physical quantities of interest for theoretical description of plasmas. In industry, these discharges can be used for plasma processing, gas discharge lamps, for the generation of chemically reactive species *etc* [2].

When these discharges operate on the right branch of the Paschen's curve, they exist in several distinct modes such as Townsend, subnormal, normal and abnormal [3]. Each of these modes has its own interesting features. Almost constant discharge voltage is typical for the Townsend and normal modes. The subnormal mode lies between these two modes and is characterized by the dropping current voltage characteristic, i.e. by the negative differential resistance.

It is known that the electrical circuits with the negative differential resistance exhibit self-sustained oscillations [4], i.e. the oscillations without an external periodic force. Self-oscillations of glow discharges in plane-to-plane geometry were studied in numerous publications (see, for instance, [5,6,7,8,9,10,11] and references therein). Different physical mechanisms responsible for the self-oscillations' onset were discussed including the relationship between the differential resistance of the discharge ($R_d$) and the external circuit resistance ($R$), and the $RC$ circuit resonance, where $C$ is either the discharge gap capacitance or the external blocking capacitance.

Due to the nonlinearity of gas discharge plasma, theoretical description of self-oscillation phenomenon is extremely difficult. Raizer *et al*. [7] have built such a model to describe the self-oscillations of current and voltage of dc discharge with semiconductor cathode. They obtained that the discharge voltage is described by the 3rd order nonlinear ordinary differential equation. They have also obtained the requirement on the differential resistance necessary for the onset of plasma self-oscillations.

Use of data-driven techniques can complement traditional theoretical approaches to get insights into the physics of complex phenomena of gas discharges. One of such approaches uses sparse regression to determine the fewest terms in the dynamic governing equations [12,13,14]. This method is called the sparse identification of nonlinear dynamics (SINDy) and is implemented in the open-source Python library PySINDy [12,13,14,15]. This library was used by Alves and Fiuza [16] to discover accurate reduced plasma models directly from the particle-in-cell simulations. They have demonstrated the recovery of fundamental hierarchy of plasma physics models from the Vlasov equation to magnetohydrodynamics.

Thakur *et al.* [17] used SINDy algorithm to "discover" the governing equations describing the anode-glow oscillations. They have obtained that these oscillations are fitted well by the 3rd order polynomials. They have also attempted to interpret the physical meaning of every term in this equation by analyzing the plasma fluid equations. They have concluded that the higher terms in the governing equation are due to the various atomic processes in plasma which they assumed to be proportional to the higher degrees of plasma density. This model does not look accurate because the probabilities of 3- and 4-body processes at the low-pressure conditions are very low.

In the present paper, the SINDy algorithm is applied to discover the governing equations of self-oscillations of dc discharge in plane-to-plane geometry. The comprehensive fluid model was used in [11] to study the plasma dynamics of this discharge. The paper is organized as follows. Section 2 briefly discusses the model of dc discharge of interest in the present studies and the data preparation for PySINDy library. Section 3 discusses the results of the present studies and presents the analytical model allowing us to interpret each term in the obtained equation. Section 4 concludes the results of the present studies.

## 2. Numerical model, data preparation and processing

In the present paper, the current-voltage characteristics of dc discharge in argon obtained in [11] are analyzed. In that paper, the discharge was modeled using the 2D axisymmetric drift-diffusion fluid model (the simulation domain geometry is shown in Figure 1(a)). The working gas was argon at the pressure of 800 Pa and the room temperature. The fluid plasma equations were coupled with the Poisson's equation for electrostatic potential. The model was also coupled with a simple circuit containing only the ballast resistor. The dc source was attached to the anode and was driven by the voltage of 500 V, while the ballast resistance was varied in the range 1 kΩ – 100 MΩ. The cathode was grounded.

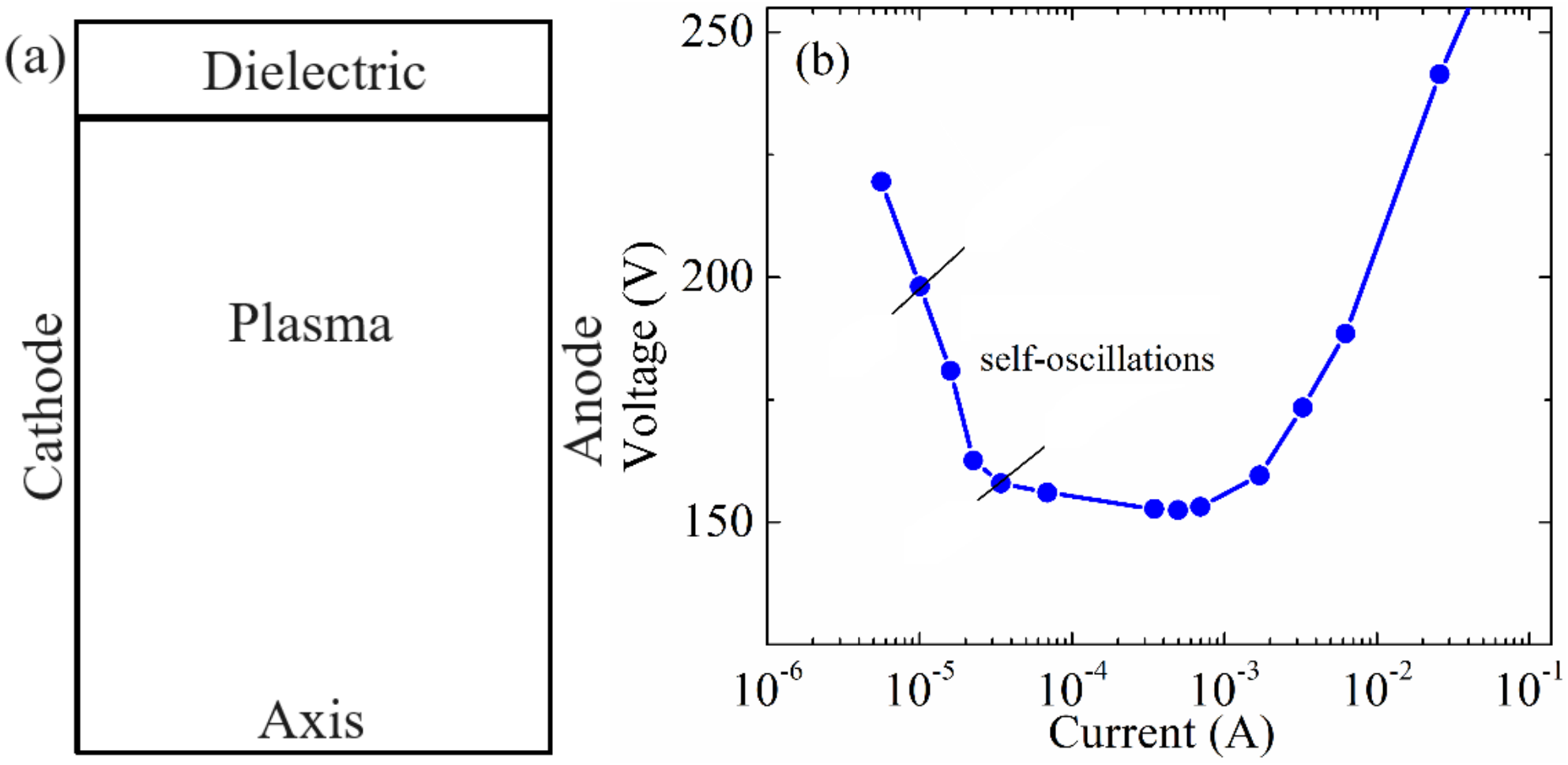

Figure 1. (a) Geometry of the simulation domain used for the modeling of direct current discharge self-oscillations. (b) Current-voltage characteristic obtained for the gas pressure of 800 Pa.

During the simulations [11], the time evolution of the anode voltage and current was saved. This data served as the "experimental" data for the data-driven discovery of governing equations with PySINDy library. The discharge current-voltage characteristic is shown in Figure 1(b). The excitation of self-oscillations was obtained in the subnormal mode of glow discharge which is characterized by the falling current-voltage characteristic, i.e. by the negative differential resistance. For the current values outside of this mode, the current-voltage oscillograms do not exhibit any oscillations.

Before applying PySINDy library, the connection voltage was first differentiated using the forward difference to determine the voltage time derivative, $dV/dt$. Then, both the voltage and $dV/dt$ were normalized by their peak values obtained during the integration time window. Once these data are obtained, PySINDy is used for the prediction of the equations describing the dynamics of connection voltage. Since the outputs of fluid simulations usually contain low numerical noise, any filtering or smoothing of this data was not applied to improve its quality.

The solution is sought in the polynomials form [12,18]:

$$\dot{\boldsymbol{X}} = \Theta(\boldsymbol{X})\Xi, \tag{1}$$

where the dot above $\boldsymbol{X}$ denotes the time derivative, $\boldsymbol{X} = [V, dV/dt]^T$, $\Theta(\boldsymbol{X})$ is a library consisting of candidate nonlinear functions of the columns of $\boldsymbol{X}$, and $\Xi$ is the vector of coefficients that determine which nonlinearities are active. In the present analysis, the polynomial library only with the polynomials order in the range 1-4 was used. For the higher order polynomials, we were not able to obtain the converged solution. The STLSQ optimizer [19] (Sequentially thresholded least squares algorithm) with the threshold value of 0.1 and maximum number of iterations of 40 has been chosen in the present analysis.

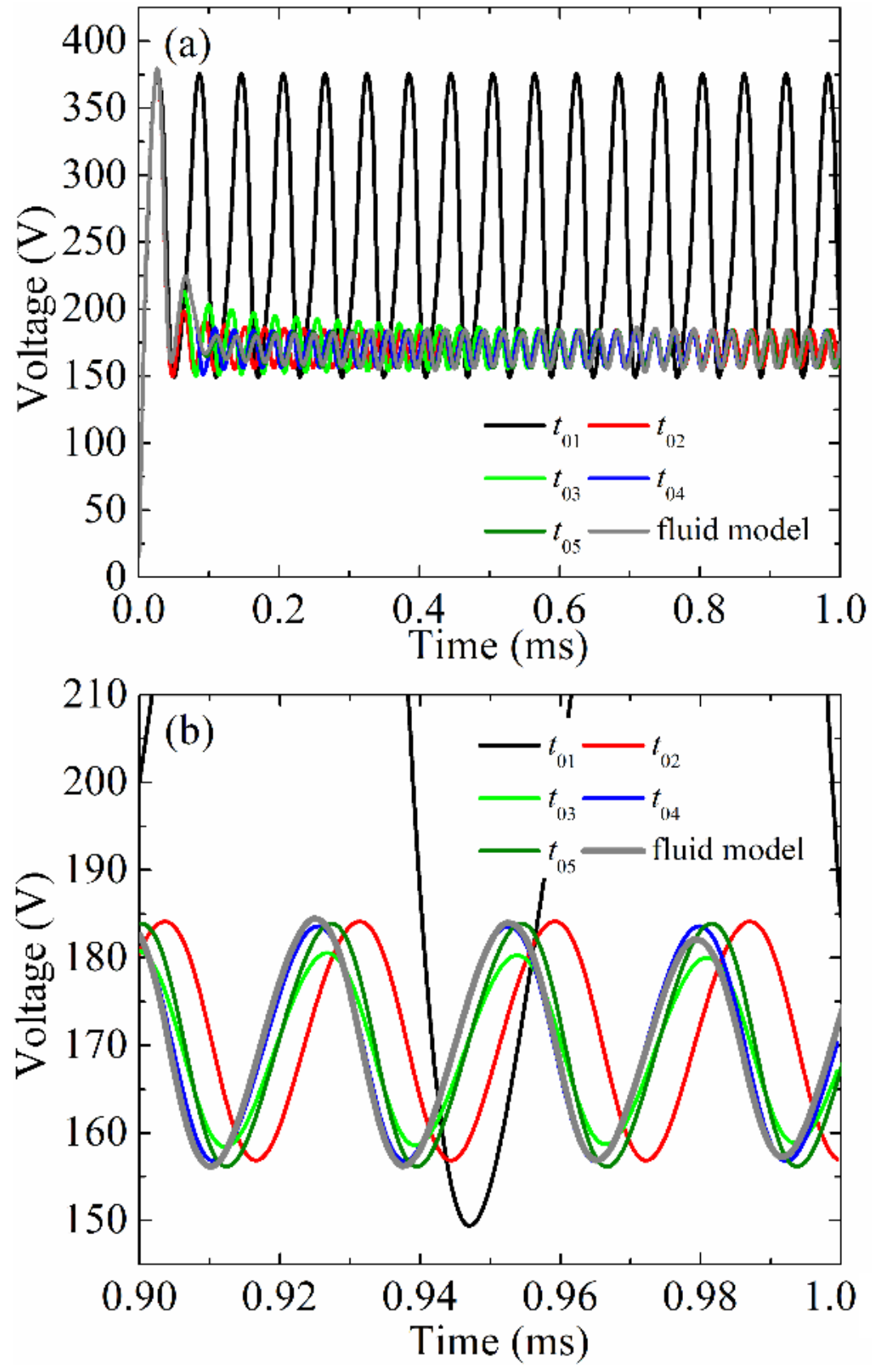

Figure 2. Voltage as the function of time obtained using fluid model and predicted by PySINDy for the 3rd order polynomials for different sets of initial conditions: (a) from the simulation start to 1 ms, and (b) from 0.9 ms to 1.0 ms.

Our analysis has shown that the initial time when PySINDy starts training on the data is crucial for an accurate prediction of governing equations. Figure 2 shows the comparison between the typical PySINDy predictions obtained for five different starting points and the fluid simulations data. One can see that if the training data captures the gap breakdown dynamics, PySINDy fails to predict the governing equations correctly describing the fluid model results. One can see that in this case the amplitude of the voltage oscillations is determined by the initial transient rather than the dynamical steady state. Also, the frequency of oscillations is smaller than that obtained from the fluid model.

As was discussed in [11], the initial voltage dynamics seen in Figure 2 at $t < 0.1$ ms is typical for both stable and unstable regimes of dc discharge. Only if certain conditions on the plasma are satisfied [3,11], the excitation of self-oscillations becomes possible. Since PySINDy does not analyze the inner plasma parameters such as the electron density and temperature but considers only the reaction in the external circuit, it is not capable of capturing these critical conditions. However, it is interesting to note that PySINDy is still able to capture the excitation of self-oscillation although having wrong amplitude and frequency.

Figure 2(b) allows us to conclude that the delay of the starting time improves the quality of predictions. One can see that for $t_{04}$ and $t_{05}$ both the voltage amplitude and the oscillations frequency match the fluid model results. Therefore, in further analysis (see Section 3) the PySINDy training starts at $t = t_{04}$. It is also important to note that for the 1$^{st}$ order polynomials, the start of training at $t = t_{01}$ results in the damping oscillator equation, while the start at $t > t_{02}$ results in the diverging oscillations.

## 3. Results and discussion

### 3.1. Data-driven discovery of the governing equations

This subsection presents the analysis for the argon gas pressure of 800 Pa, the cathode-anode gap distance $d = 1$ cm, the source voltage of 500 V and the ballast resistance of $R = 20$ MΩ. For this value of $R$, the discharge self-oscillations were obtained. Similar results were obtained for other values of $R$ for which the self-oscillations were excited but with other coefficients in the discovered equations (see discussion in Section 3.2).

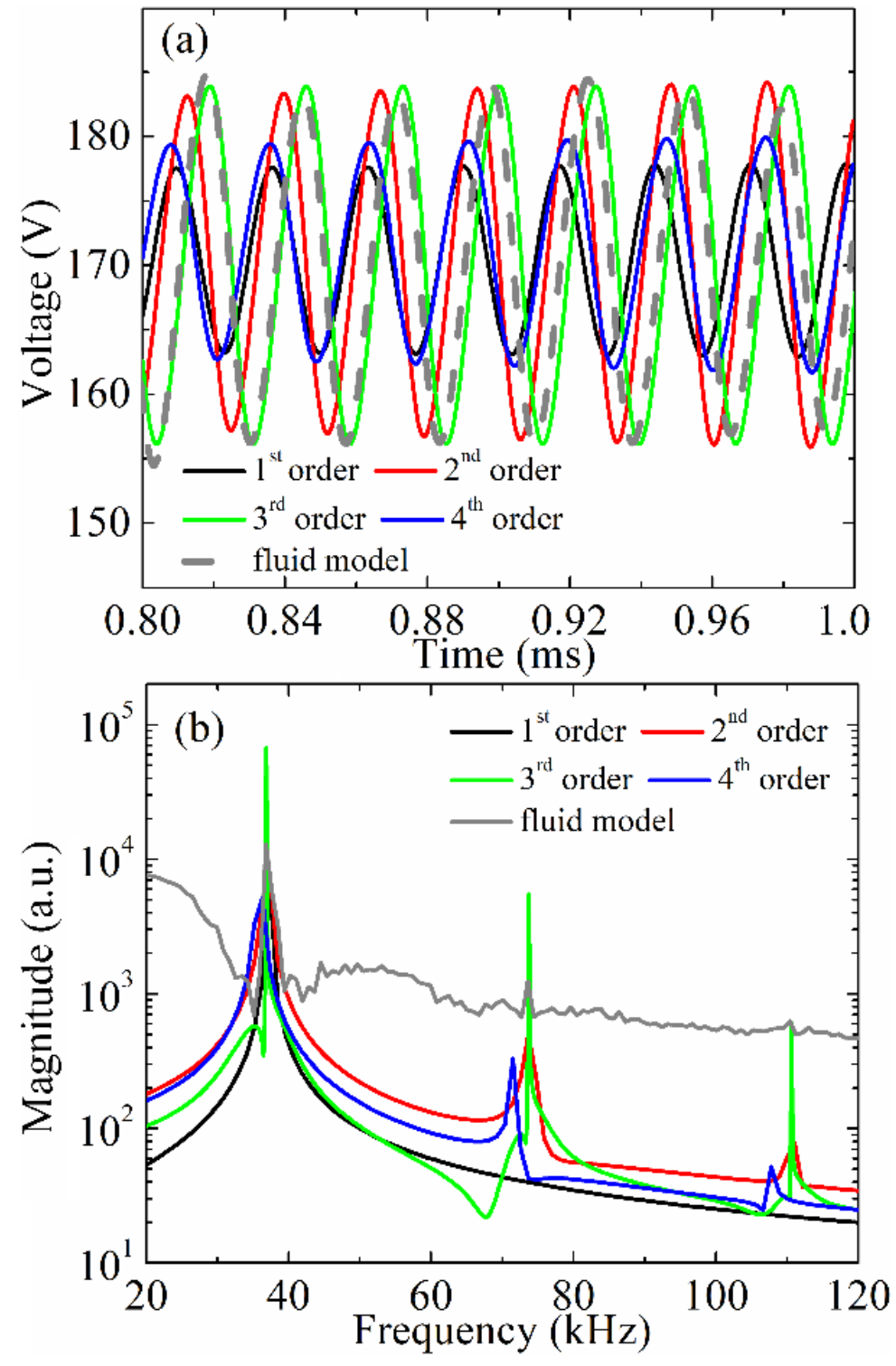

Figure 3. (a) Time evolution of the discharge voltage obtained using the fluid model and PySINDy for different polynomial orders. (b) Fourier spectra obtained from the fluid model and PySINDy for different polynomial orders.

Let us use the following notation:

$$\begin{bmatrix} x_0 \\ x_1 \end{bmatrix} = \begin{bmatrix} V \\ dV/dt \end{bmatrix}.$$

Then, PySINDy library is used to obtain the coefficients in the system of equations [12]:

$$\begin{cases} \dot{x}_0 = a_1 x_0^0 + b_1 x_0^1 + c_1 x_1^1 + d_1 x_0^2 + e_1 x_0^1 x_1^1 + f_1 x_1^2 + g_1 x_0^3 + h_1 x_0^2 x_1^1 + k_1 x_0^1 x_1^2 + l_1 x_1^3, \\ \dot{x}_1 = a_2 x_0^0 + b_2 x_0^1 + c_2 x_1^1 + d_2 x_0^2 + e_2 x_0^1 x_1^1 + f_2 x_1^2 + g_2 x_0^3 + h_2 x_0^2 x_1^1 + k_2 x_0^1 x_1^2 + l_2 x_1^3 \end{cases} \tag{2}$$

which fits best the voltage time evolution obtained from the fluid simulations. Note that equation (2) shows the 3rd order polynomials only for the illustration; 1st, 2nd and 4th order polynomials were considered as well.

**TABLE 1. Coefficients in equations (2) derived with PySINDy for the 2nd and 3rd order polynomials for the gas pressure of 800 Pa, the source voltage of 500 V and the ballast resistance of $R$ = 20 MΩ.**

| | 2nd order | | 3rd order | |
|---|---|---|---|---|
| | $x_0'$ | $x_1'$ | $x_0'$ | $x_1'$ |
| $x_0^0$ $(a)$ | -14684.555 | 3082792.009 | -214837.473 | 39514055.455 |
| $x_0^1$ $(b)$ | 18186.765 | -3593863.033 | 681737.344 | -124872169.177 |
| $x_1^1$ $(c)$ | 23883.912 | -1455413.140 | 18255.482 | -2285752.918 |
| $x_0^2$ $(d)$ | -2412.706 | 256309.910 | -733710.571 | 134473154.461 |
| $x_0^1 x_1^1$ $(e)$ | -6850.282 | 1574395.367 | 4933.468 | 3541147.421 |
| $x_1^2$ $(f)$ | 129.275 | -15440.476 | -2491.838 | 481685.890 |
| $x_0^3$ $(g)$ | - | - | 267928.671 | -49375397.385 |
| $x_0^2 x_1^1$ $(h)$ | - | - | -6163.576 | -1137185.060 |
| $x_0^1 x_1^2$ $(k)$ | - | - | 2897.869 | -560971.010 |
| $x_1^3$ $(l)$ | - | - | 36.556 | -26081.457 |

The PySINDy prediction for different polynomial orders is shown in Figure 3(a). It is seen that the best agreement is obtained for the 3rd order polynomials, although the 2nd order polynomials can also be used for a qualitative description of the discharge voltage dynamics. The difference in physical mechanisms behind the 2nd and the 3rd order polynomials is discussed in Section 3.2. In both cases, the discovered equations predict the voltage amplitude rather well. However, the 2nd order polynomials predict slight phase shift compared with the fluid model results, while the 3rd order polynomials match the fluid model better. It is important to note that both governing equations predict the oscillation frequency rather well (see the Fourier spectra in Figure 3(b)). They predict not only the location of the fundamental harmonic but also accurately predict the location of the second and the third harmonics. The coefficients of equation (2) for the 2nd and the 3rd order polynomials are shown in Table 1.

Figure 3(a) shows that the 1st and the 4th order polynomials cannot predict the discharge voltage dynamics. One can see that on the time scale of 1 ms both predict smaller voltage amplitude, and the time lag compared to the fluid model results. However, Figure 3(b) shows that both the 1$^{st}$ and the 4$^{th}$ order polynomials predict the location of the fundamental harmonic rather well. Note that since the 1$^{st}$ order equation describes the harmonic oscillator with an external force, its solution does not contain higher harmonics. The 4$^{th}$ order equation predicts higher harmonics having smaller values.

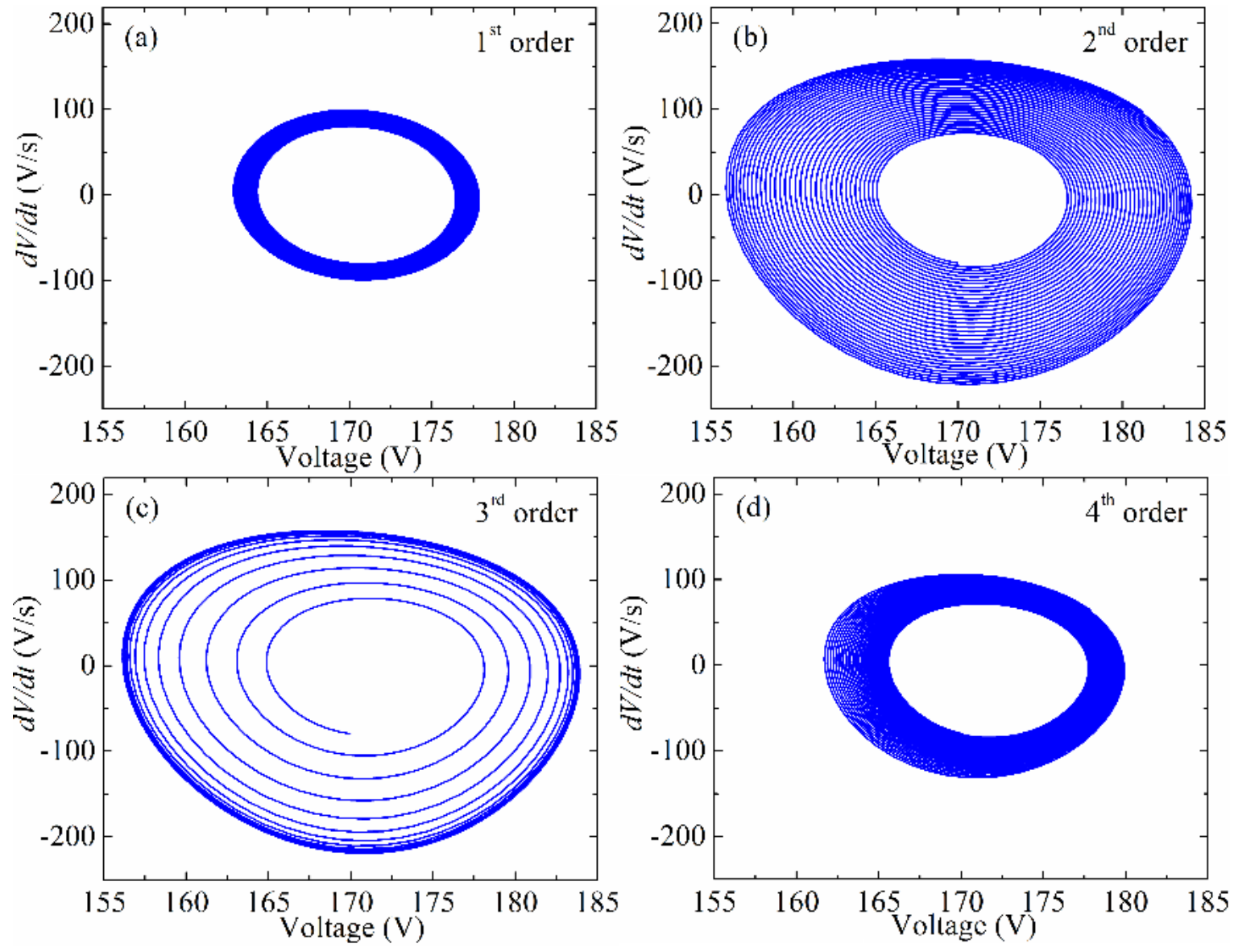

Figure 4. Phase trajectories obtained using PySINDy for different polynomials: (a) 1$^{st}$ order, (b) 2$^{nd}$ order, (c) 3$^{rd}$ order, and (d) 4$^{th}$ order.

The phase trajectories of the obtained equations are shown in Figure 4. One can see in Figure 4(b,c) that the solution of both the 2$^{nd}$ and the 3$^{rd}$ order polynomial equations eventually converge to the same attractor. The phase trajectories of the 1$^{st}$ and the 4$^{th}$ order equations are shown in Figure 4(a,d), respectively. One can see that these two equations establish different behavior. The 1$^{st}$ order equation describes the diverging oscillations, i.e. covers the entire phase space at $t \to \infty$. The solution of the 4$^{th}$ order equation has an attractor with a radius smaller than that of the 2$^{nd}$ and the 3$^{rd}$ order equations.

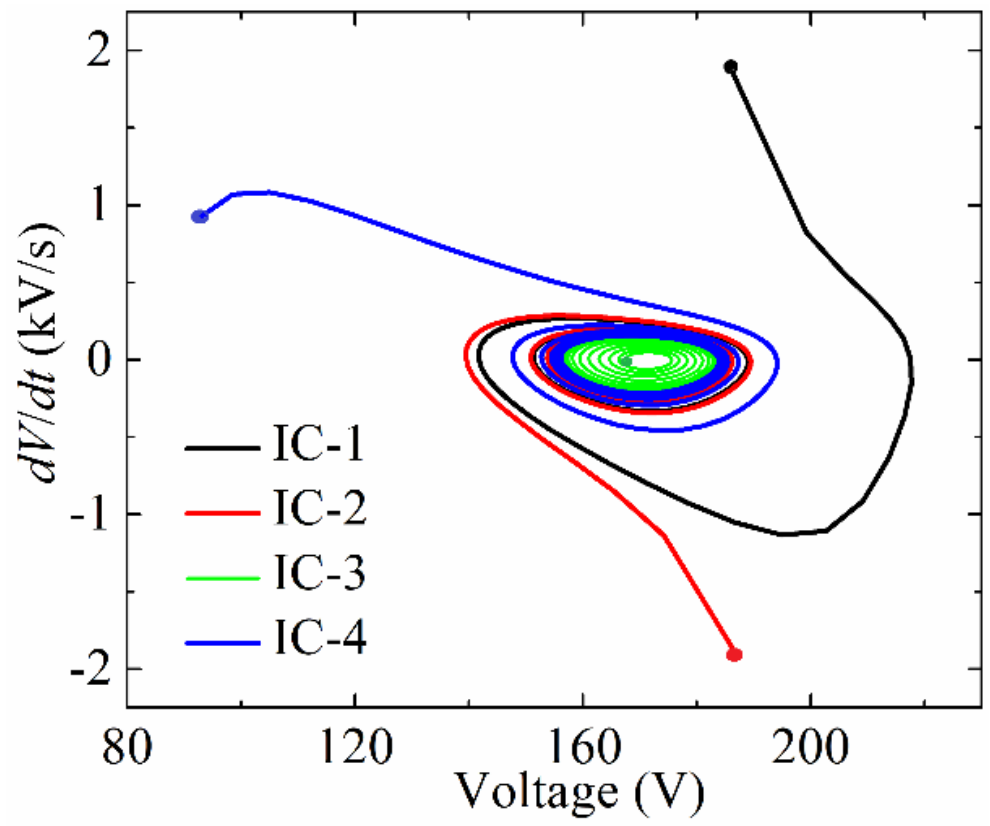

Figure 5. Limit-cycle behavior of the 3$^{rd}$ order model equation for four sets of initial conditions (IC) shown by cycles.

The solutions of the 2nd and the 3rd order equations display the stable limit cycle behavior. Figure 5 demonstrates this for the 3rd order equation; the 2nd order equation exhibits similar behavior. The existence of this limit cycle supports our conclusion that the discovered model equations exhibit self-sustained oscillations, i.e. the system oscillates even in the absence of the external periodic forcing [20]. Figure 5 shows the solutions for four different initial conditions, one of which is inside the attractor and three others are outside. One can see that independently of where the trajectory starts it converges to the same attractor.

### 3.2. Theoretical model of voltage self-oscillations

In this subsection, we consider the one-dimensional model of dc discharge to give the physical meaning of every term in equation (2). This model is based on the one developed in [7] where Raizer *et al.* considered the cathode made of semiconductor. They also accounted for the electron secondary emission from the cathode due to the metastable species impact. In our fluid model of dc discharge [11], these effects were not considered.

Consider the Townsend discharge, i.e. ignore the influence of the plasma space charge on the electric field distribution. Thus, the electric field is homogeneous across the cathode-anode gap. The discharge is supported by the secondary electron emission from the cathode due to the ion impact with constant secondary emission coefficient $\gamma$ as in the numerical model of [11].

The equivalent circuit of the discharge gap consists of the capacitance $C_p$ connected in parallel with the resistance $R_p$ (see Figure 6). The latter is due to the electron and ion conduction currents through the gap. Then, the total voltage is defined as

$$V_0 = IR + V, \tag{3}$$

where $I$ and $V$ are the discharge current and voltage, respectively. The plasma current consists of both the conduction and the displacement currents. Then, one can define the discharge current as

$$I = S\left(j + \frac{\varepsilon_0}{d}\frac{dV}{dt}\right), \tag{4}$$

where $S$ is the electrode surface area and $\varepsilon_0$ is the permittivity of a free space. Thus, one obtains the equation for the cathode-anode gap voltage:

$$\frac{dV}{dt} = \frac{V_0 - V - SRj}{\theta}, \tag{5}$$

where $\theta = RC$ is the discharge time constant.

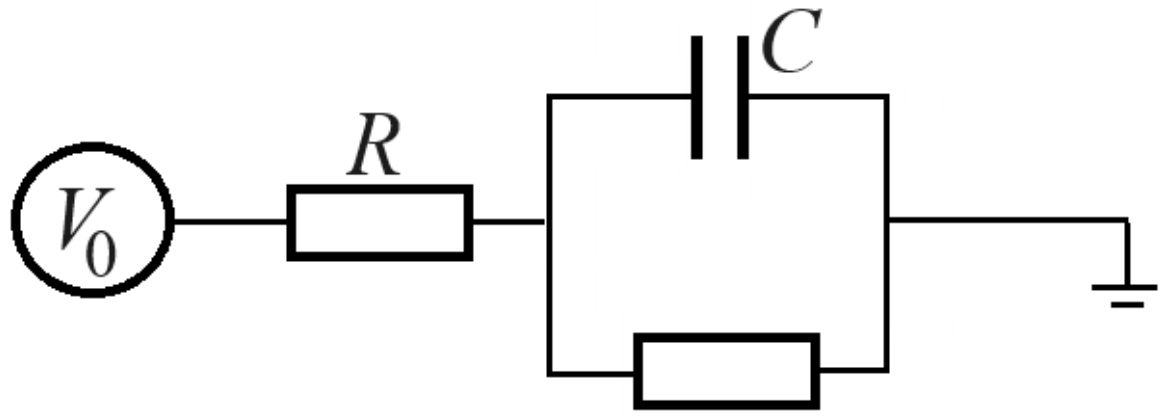

Figure 6. Equivalent scheme of the Townsend discharge considered in the present paper.

In the steady state of Townsend discharge the following condition must be fulfilled [21]:

$$M(t) = \gamma\left\{exp\left[\int_0^d \alpha(x,t)dx\right] - 1\right\} = 1, \quad (6)$$

where $\alpha(x,t)$ is the Townsend ionization coefficient as the function of space and time. The condition $M(t) = 1$ means that the electron losses to the walls are compensated by the multiplication of the secondary emitted electrons. The expression in the brackets defines the ion density (current) generated due to the multiplication of secondary emitted electrons in the cathode-anode gap.

Following Raizer *et al*. [7], equation (6) can be re-written as

$$j_e(t+\tau) = \gamma\left\{exp\left[\int_0^d \alpha(x,t)dx\right] - 1\right\} j_e(t), \quad (7)$$

where $j_e$ is the electron current from the cathode due to the secondary emission and $\tau$ is the ion transit time through the cathode-anode gap. This equation considers that there is the time delay between the ion current generated by the multiplication of the secondary emitted electrons and the arrival of these ions to the cathode.

The discharge current density at the anode consists of two components, the current of secondary emitted electrons and the current of electrons generated due to the gas ionization

$$j = j_e + \frac{j_e}{\gamma}, \quad (7)$$

which allows to define the electron current as

$$j_e = \frac{\gamma}{\gamma+1} j. \quad (8)$$

Substituting this equation into (7) and then expanding $j(t+\tau)$ into the Taylor series around $\tau$, one obtains the equation for the discharge current density:

$$\frac{dj}{dt} = \frac{g-1}{\tau} j, \quad (9)$$

where $g = \gamma\left(exp\left[\int_0^d \alpha(x,t)dx\right] - 1\right)$. This equation is like that obtained in [7]. However, equation (9) does not consider the time dependence of the secondary electron emission coefficient and neglects the small constant electron current from the cathode. As was pointed out in [7], this current may be due to the secondary electron emission from the cathode caused by the metastable atoms' bombardment. This mechanism was absent in our numerical fluid model of [11]. Therefore, it is not considered in deriving (9) as well.

Thus, we have the system of two equations for the discharge voltage (5) and the discharge current density (9). In (9), the Townsend ionization coefficient depends on the electric field which can be approximately defined as $E \approx -\frac{V}{d}$. Then, the multiplication coefficient can be expanded around the current-voltage characteristic $V = V_s(j)$ [7]:

$$g - 1 = \left(\frac{dg}{dV}\right)_{V=V_s(j)} (V - V_s(j)). \quad (10)$$

The Townsend ionization coefficient is usually approximated as $\alpha = Ap\exp(-Bp/E)$ [21], where $A$ and $B$ are empirical constants. In the homogeneous electric field, it is defined as $\alpha = Ap\exp(-Bpd/V)$. To estimate $\left(\frac{dg}{dV}\right)$ in (10), one can use the condition $M(t) = 1$ which should be satisfied in the stationary regime of discharge. Then, using (7) to obtain $\int_0^d \alpha(x,t)dx$, one gets:

$$\left(\frac{dg}{dV}\right)_{V=V_s(j)} = (1+\gamma)\ln(1+1/\gamma)\left(\frac{d\alpha}{dV}\right)_{V=V_s(j)}. \tag{11}$$

In this equation, $\alpha$ is the known function of the discharge voltage. Now, substituting equations (10) and (11) into (9), one obtains the final equation for the discharge current density [7]:

$$\frac{dj}{dt} = \frac{k}{\tau_0}\frac{V-V_s(j)}{V_s(j)}j. \tag{12}$$

Here, $k = \frac{(1+\gamma)L_\gamma}{1-L_\gamma^{-1}}$, $L_\gamma = \ln(1+1/\gamma)$ and $\tau_0 = \frac{d}{\mu_i E_0}$ is the ion transit time through the cathode-anode gap in the stationary regime in the homogeneous electric field $E_0$.

The system of equations (5) and (12) should be closed by the current-voltage characteristics. Raizer *et al.* [7] have derived the following expression for the subnormal mode of the glow discharge:

$$V_s(j) = V_T - R_d j + A_1 j^2. \tag{13}$$

Here, $V_T$ is the breakdown voltage and $A_1 = -\frac{1}{24V_T^3}\left(\frac{Bpd}{V_T} - 2\right)\frac{d^5}{\varepsilon_0\mu_i^2}$ is constant [7]. It depends on the derivatives of the coefficient $\alpha$ and the breakdown voltage. In the low current mode, the discharge voltage differs insignificantly from the breakdown voltage. Therefore, it is possible to assume that $V_s(j) \approx V_T$ in the denominator of (12) [7]. Then, using equation (13), one gets from (5) and (12) the nonlinear differential equation describing the dynamics of the discharge voltage:

$$\frac{V_T\tau_0\theta}{k}\ddot{V} = a + bV + c\dot{V} + dV^2 + eV\dot{V} + f\dot{V}^2 + gV^3 + hV^2\dot{V} + kV\dot{V}^2 + l\dot{V}^3. \tag{14}$$

This equation coincides with the second of equations (2) for $\dot{x}_1 = \ddot{V}$. The 3rd order terms in this equation are due to the quadratic term in the current-voltage characteristic (13). The coefficients in equation (14) are defined as:

$$a = V_0\left(V_T + \frac{A_1V_0^2}{(SR)^2} - \frac{R_d}{R}V_0\right),\ b = \frac{2R_d}{R}V_0 - V_0 - V_T - \frac{3A_1V_0^2}{(SR)^2},\ c = \frac{2R_d\theta}{R}V_0 - \theta V_T - \frac{V_T\tau_0}{k} - \frac{3A_1\theta V_0^2}{(SR)^2},\ d = 1 - \frac{R_d}{R} + \frac{3A_1V_0}{(SR)^2},\ e = \frac{4A_1\theta V_0}{(SR)^2} - \frac{R_d}{R}\theta,\ f = \theta^2\left(\frac{3A_1V_0}{(SR)^2} - \frac{R_d}{R}\right),\ g = \frac{A_1}{(SR)^2},\ h = \frac{3A_1\theta}{(SR)^2},\ k = \frac{3A_1\theta}{(SR)^2},\ l = \frac{A_1\theta^3}{(SR)^2}. \tag{15}$$

Here, it is important to note that similar the 3rd order ordinary different equation was derived in [7] for the plane-to-plane geometry with the semiconductor cathode.

Equation (14) can be re-written as follows:

$$\ddot{V} - \frac{k}{V_T\tau_0\theta}(c + hV^2)\dot{V} + \frac{k}{V_T\tau_0\theta}(-b - gV^2 - k\dot{V}^2)V = \frac{k}{V_T\tau_0\theta}\left(a + dV^2 + eV\dot{V} + f\dot{V}^2 + l\dot{V}^3\right). \tag{16}$$

This is the van der Pol equation [20] with the modulated frequency and the right-hand side which depends on the discharge voltage. This equation allows the excitation of self-oscillations if the term $\frac{kc}{V_T \tau_0 \theta}$ is positive. This gives the condition on the discharge differential resistance:

$$\frac{R_d}{R} > \frac{1}{2}\left(\frac{V_T}{V_0}\left(1+\frac{\tau_0}{k\theta}\right)+\frac{3A_1 V_0}{(SR)^2}\right). \tag{17}$$

Note that since $A_1$ is the negative parameter depending on $V_T$, the term $\frac{kc}{V_T \tau_0 \theta}$ can also be negative, i.e. there is no excitation of self-oscillations. This is obtained if the differential resistance does not satisfy the condition (17).

It was obtained in Section 3.1 that the 2nd order differential equation also describes the dynamics of the discharge voltage rather well. As it follows from the current-voltage characteristic (13), the 2nd order equation is obtained when the voltage is the linear function of the current density. Thus, equation (16) can be simplified as follows:

$$\ddot{V} - \frac{kc}{V_T \tau_0 \theta}\dot{V} + \frac{-bk}{V_T \tau_0 \theta}V = \frac{k}{V_T \tau_0 \theta}\left(a + dV^2 + eV\dot{V} + f\dot{V}^2\right). \tag{18}$$

Note that the coefficients $a$, $b$, $c$, $d$, $e$ and $f$ of the 2nd order equation differ from those of the 3rd order equation (see Table 1). As at follows from the results shown in Figure 3 and Figure 4 this 2nd order ordinary differential equation allows the excitation of self-oscillations as well. Comparison between the Fourier spectra of equations (16) and (18) shown in Figure 3(b) allows us to conclude that the 3rd order terms define the phase shift, while they have weak influence on the frequencies and amplitude of oscillations. The terms proportional to $V^2$ and $\dot{V}^2$ do not change their signs during the period and, consequently, define the higher harmonics.

### 3.3. Influence of the ballast resistance on the coefficients in equation (2)

In this subsection, we analyze the coefficients in equation (2) for various values of the ballast resistance. Large values of $R$ correspond to the low current, i.e. Townsend, mode, while small values of $R$ correspond to the normal mode (see Figure 1(b)). Discharge self-oscillations were obtained for $R =$ 18-30 MΩ. Our analysis has shown that the 3rd order differential equation (2) describes rather well the time evolution of the discharge voltage even outside the region of self-oscillations existence.

Table 2 shows the values of the coefficients in the second of equations (2) which form coincides with equation (16). It is important to note that all coefficients are non-zero even in the region where there is no self-oscillations. One can see that for $R =$ 18-30 MΩ coefficient $c$ defined as $c = \frac{2R_d\theta}{R}V_0 - \theta V_T - \frac{V_T \tau_0}{k} - \frac{3A_1\theta V_0^2}{(SR)^2}$ (see equation (15)) is negative which means that the conditions (17) on the differential resistance is fulfilled.

**TABLE 2. Coefficients of the 3[rd] order differential equation (2) for various values of the ballast resistance.**

| *R*, MΩ | *a* | *b* | *c* | *d* | *e* | *f* | *g* | *h* | *k* | *l* |
|---|---|---|---|---|---|---|---|---|---|---|
| 14 | $-1.0743\times10^{11}$ | $3.2563\times10^{11}$ | $9.0436\times10^{8}$ | $-3.2897\times10^{11}$ | $-1.8194\times10^{9}$ | $-1.5615\times10^{7}$ | $1.1077\times10^{11}$ | $9.1501\times10^{8}$ | $1.6093\times10^{7}$ | $-3.9886\times10^{5}$ |
| 16 | $4.5368\times10^{9}$ | $-1.4019\times10^{10}$ | $5.2113\times10^{7}$ | $1.4449\times10^{10}$ | $-1.0805\times10^{8}$ | $-9.6567\times10^{5}$ | $-4.9675\times10^{9}$ | $5.5903\times10^{7}$ | $1.0227\times10^{6}$ | $3.0083\times10^{4}$ |
| 18 | $1.0753\times10^{8}$ | $-2.9089\times10^{8}$ | $-2.2445\times10^{7}$ | $2.6049\times10^{8}$ | $4.5834\times10^{7}$ | $1.1464\times10^{5}$ | $-7.7355\times10^{7}$ | $-2.337\times10^{7}$ | $-7.0540\times10^{4}$ | $-3.5374\times10^{4}$ |
| 20 | $3.9514\times10^{7}$ | $-1.2487\times10^{8}$ | $-2.2858\times10^{6}$ | $1.3447\times10^{8}$ | $3.5411\times10^{6}$ | $4.8169\times10^{5}$ | $-4.9375\times10^{7}$ | $-1.1372\times10^{6}$ | $-5.6097\times10^{5}$ | $-2.6081\times10^{4}$ |
| 22 | $1.0454\times10^{7}$ | $-3.1937\times10^{7}$ | $-2.1075\times10^{6}$ | $3.3616\times10^{7}$ | $3.7292\times10^{6}$ | $2.6625\times10^{5}$ | $-1.2356\times10^{7}$ | $-1.4796\times10^{6}$ | $-3.3302\times10^{5}$ | $-6.9404\times10^{3}$ |
| 24 | $1.0201\times10^{7}$ | $-3.4231\times10^{7}$ | $-2.0732\times10^{6}$ | $3.9288\times10^{7}$ | $4.0694\times10^{6}$ | $3.3655\times10^{5}$ | $-1.5456\times10^{7}$ | $-1.8956\times10^{6}$ | $-4.5333\times10^{5}$ | $-5.4022\times10^{3}$ |
| 26 | $9.2542\times10^{7}$ | $-3.2032\times10^{7}$ | $-1.9153\times10^{6}$ | $3.7706\times10^{7}$ | $3.9205\times10^{6}$ | $3.1691\times10^{5}$ | $-1.5130\times10^{7}$ | $-1.9266\times10^{6}$ | $-4.2396\times10^{5}$ | $4.7564\times10^{3}$ |
| 28 | $7.9096\times10^{6}$ | $-2.7372\times10^{7}$ | $-1.9071\times10^{6}$ | $3.2331\times10^{7}$ | $4.0851\times10^{6}$ | $2.8632\times10^{5}$ | $-1.3035\times10^{7}$ | $-2.1359\times10^{6}$ | $-4.019\times10^{5}$ | $-2.1603\times10^{3}$ |
| 30 | $5.9946\times10^{6}$ | $-2.0302\times10^{7}$ | $-2.0756\times10^{6}$ | $2.3601\times10^{7}$ | $4.7392\times10^{6}$ | $2.0477\times10^{5}$ | $-9.4663\times10^{6}$ | $-2.6764\times10^{6}$ | $-2.88\times10^{5}$ | $2.7329\times10^{3}$ |
| 34 | $7.3339\times10^{7}$ | $-2.1849\times10^{8}$ | $-1.4240\times10^{7}$ | $2.1587\times10^{8}$ | $3.1143\times10^{7}$ | $5.8369\times10^{5}$ | $-7.0908\times10^{7}$ | $-1.7017\times10^{7}$ | $-5.6722\times10^{5}$ | $1.6212\times10^{4}$ |
| 40 | $6.2456\times10^{7}$ | $-1.9773\times10^{8}$ | $-8.8126\times10^{6}$ | $2.0888\times10^{8}$ | $1.8961\times10^{7}$ | $1.1622\times10^{6}$ | $-7.3685\times10^{7}$ | $-1.0171\times10^{7}$ | $-1.1432\times10^{6}$ | $-1.3074\times10^{5}$ |
| 50 | $4.0509\times10^{7}$ | $-1.2437\times10^{8}$ | $-1.5711\times10^{7}$ | $1.2773\times10^{8}$ | $3.2681\times10^{7}$ | $1.6907\times10^{6}$ | $-4.3928\times10^{7}$ | $-1.6995\times10^{7}$ | $-1.7125\times10^{6}$ | $-1.2137\times10^{5}$ |
| 60 | $1.4112\times10^{7}$ | $-4.5443\times10^{7}$ | $-1.6208\times10^{6}$ | $4.9246\times10^{7}$ | $3.0979\times10^{6}$ | $5.3307\times10^{5}$ | $-1.7976\times10^{7}$ | $-1.4875\times10^{6}$ | $-5.2984\times10^{5}$ | $-3.5719\times10^{4}$ |

One can conclude from Table 2 that in the transition region from the sub-normal to normal mode (region between 16 and 18 MΩ) coefficient $c$ changes its sign. Therefore, the condition (16) is not satisfied for smaller resistance / higher discharge current. Indeed, Figure 1(b) shows that the normal mode is still characterized by the negative slope of the current-voltage characteristic. However, the differential resistance obtained for this mode is much smaller than that of the sub-normal mode. Table 2 also shows that the values of all coefficients increase by ~1-2 orders of magnitude.

The sign of coefficient $c$ does not change when transiting from the sub-normal to Townsend mode ($R > 30$ MΩ). Moreover, none of the other coefficients changes its sign. However, as in the case of the sub-normal-to-normal mode transition, all the coefficients change drastically, namely, they increase by ~5-20 times. The non-linear damping term $(c + hV^2)\dot{V}$ in equation (16) defines the character of solution. Our analysis has shown that even a small increase of either $c$ or $h$, or both results in the damping of oscillations. This, probably, explains why the increase of the ballast resistance / decrease of the discharge current leads to the damping of self-oscillations obtained in our studies.

## 4. Summary

In the present paper, the current-voltage characteristic of the direct current discharge in argon obtained from the self-consistent two-dimensional plasma model was analyzed using the sparse identification of nonlinear dynamics (SINDy) method. The discharge operated in the subnormal mode which is characterized by the excitation of self-sustaining oscillations.

We have obtained with the use of the open-source library PySINDy that the voltage self-oscillations are described reasonably well by the 2nd and the 3rd order nonlinear ordinary differential equations. These equations predicted both the amplitude of the voltage oscillations and their frequency. They also predicted correct frequencies of the higher harmonics. The analysis has shown that the 4th order equation does not describe the discharge voltage dynamics, namely, the predicted voltage amplitude is much smaller than that obtained from the fluid simulations. Also, the frequencies of higher harmonics are smaller.

To understand the physical meaning of each term, an analytical model was presented which describes the oscillations of discharge voltage. This model supports the equations discovered with the data-driven approach. It is concluded that the 2nd order terms are obtained from the linear terms in the discharge current-voltage characteristics, while the 3rd order terms are due to the quadratic dependence of the discharge voltage on current.

**Data availability statement**

The data that support the finding of this study are available from the author upon reasonable request.

---